\global\newcount\numsec\global\newcount\numfor
\gdef\profonditastruttura{\dp\strutbox}

\def\senondefinito#1{\expandafter\ifx\csname#1\endcsname\relax}

\def\SIA #1,#2,#3 {\senondefinito{#1#2}%
\expandafter\xdef\csname #1#2\endcsname{#3}\else
\write16{???? ma #1,#2 e' gia' stato definito !!!!} \fi}

\def\etichetta(#1){(\veroparagrafo.\veraformula)%
\SIA e,#1,(\veroparagrafo.\veraformula) %
\global\advance\numfor by 1%
\write15{\string\FU (#1){\equ(#1)}}%
\write16{ EQ \equ(#1) <==#1  }}

\def\FU(#1)#2{\SIA fu,#1,#2 }

\def\etichettaa(#1){(A\veroparagrafo.\veraformula)%
\SIA e,#1,(A\veroparagrafo.\veraformula) %
\global\advance\numfor by 1%
\write15{\string\FU (#1){\equ(#1)}}%
\write16{ EQ \equ(#1) <== #1  }}

\def\getichetta(#1){Fig. \verafigura
\SIA e,#1,{\verafigura} %
\global\advance\numfig by 1%
\write15{\string\FU (#1){\equ(#1)}}%
\write16{ Fig. \equ(#1) ha simbolo  #1  }}

\newdimen\gwidth

\def\BOZZA{
\def\alato(##1){%
 {\vtop to \profonditastruttura{\baselineskip
 \profonditastruttura\vss
 \rlap{\kern-\hsize\kern-1.2truecm{$\scriptstyle##1$}}}}}
\def\galato(##1){\gwidth=\hsize \divide\gwidth by 2%
 {\vtop to \profonditastruttura{\baselineskip
 \profonditastruttura\vss
 \rlap{\kern-\gwidth\kern-1.2truecm{$\scriptstyle##1$}}}}}
}

\def\alato(#1){}
\def\galato(#1){}

\def\veroparagrafo{\number\numsec}\def\veraformula{\number\numfor}
\def\verafigura{\number\numfig}

\def\geq(#1){\getichetta(#1)\galato(#1)}
\def\Eq(#1){\eqno{\etichetta(#1)\alato(#1)}}
\def\eq(#1){\etichetta(#1)\alato(#1)}
\def\Eqa(#1){\eqno{\etichettaa(#1)\alato(#1)}}
\def\eqa(#1){\etichettaa(#1)\alato(#1)}
\def\eqv(#1){\senondefinito{fu#1} #1
\write16{#1 non e' (ancora) definito}%
\else\csname fu#1\endcsname\fi}
\def\equ(#1){\senondefinito{e#1}\eqv(#1)\else\csname e#1\endcsname\fi}

\def\include#1{
\openin13=#1.aux \ifeof13 \relax \else
\input #1.aux \closein13 \fi}

\openin14=\jobname.aux \ifeof14 \relax \else
\input \jobname.aux \closein14 \fi
\openout15=\jobname.aux
\def\e{\epsilon}

\def \n{\nabla}


\overfullrule=0pt
\magnification=\magstep1
\baselineskip=12pt
\outer\def\beginsection#1\par{\vskip0pt plus.3\vsize\penalty-100
   \vskip0pt plus-.3\vsize\bigskip\vskip\parskip
   \message{#1}\leftline{\bf#1}\nobreak\smallskip}
\def\IR{{\rm I\kern -1.6pt{\rm R}}}
\def\IP{{\rm I\kern -1.6pt{\rm P}}}
\def\ZZ{{\rm Z\kern -4.0pt{\rm Z}}}
\def\IC{\ {\rm I\kern -6.0pt{\rm C}}}

\def\o{\omega}

\let\hat=\widehat
\def\sqr#1#2{{\vcenter{\vbox{\hrule height.#2pt
\hbox{\vrule width.#2pt height #1pt \kern#1pt
\vrule width.#2pt}
\hrule height.#2pt}}}}
\def\square{\mathchoice\sqr56\sqr56\sqr{2.1}3\sqr{1.5}3}

\def\pmb#1{\setbox0=\hbox{$#1$}%
\kern-.025em\copy0\kern-\wd0
\kern.05em\copy0\kern-\wd0
\kern-.025em\raise.0433em\box0 }
 
\def\pmbb#1{\setbox0=\hbox{$\scriptstyle#1$}%
\kern-.025em\copy0\kern-\wd0
\kern.05em\copy0\kern-\wd0
\kern-.025em\raise.0433em\box0 }

\def\mfr#1/#2{\hbox{${{#1} \over {#2}}$}}
\catcode`@=11
\def\eqalignii#1{\,\vcenter{\openup1\jot \m@th
\ialign{\strut\hfil$\displaystyle{##}$&
        $\displaystyle{{}##}$\hfil&
        $\displaystyle{{}##}$\hfil\crcr#1\crcr}}\,}
\catcode`@=12
\def\uprho{\raise1pt\hbox{$\rho$}}


\catcode`@=11
\def\eqalignii#1{\,\vcenter{\openup1\jot \m@th
\ialign{\strut\hfil$\displaystyle{##}$& $\displaystyle{{}##}$\hfil&
$\displaystyle{{}##}$\hfil\crcr#1\crcr}}\,} \catcode`@=12

\def\R{{\rm I\kern -1.6pt{\rm R}}} 
\def\go{\mathrel{\rlap{\lower.6ex\hbox{$\sim$}}\raise.35ex\hbox{$>$}}}
\def\lo{\mathrel{\rlap{\lower.6ex\hbox{$\sim$}}\raise.35ex\hbox{$<$}}}
\def\boxit#1{\thinspace\hbox{\vrule\vtop{\vbox{\hrule\kern1pt
\hbox{\vphantom{\tt/}\thinspace{\tt#1}\thinspace}}\kern1pt\hrule}\vrule}
\thinspace}
\def\d{{\rm d}^3v}
\def\vt{\langle |v|^2 \rangle_t}
\def\vf{\langle |v|^4 \rangle_t}
%
%
%
\catcode`\X=12\catcode`\@=11
\def\n@wcount{\alloc@0\count\countdef\insc@unt}
\def\n@wwrite{\alloc@7\write\chardef\sixt@@n}
\def\n@wread{\alloc@6\read\chardef\sixt@@n}
\def\crossrefs#1{\ifx\alltgs#1\let\tr@ce=\alltgs\else\def\tr@ce{#1,}\fi
   \n@wwrite\cit@tionsout\openout\cit@tionsout=\jobname.cit 
   \write\cit@tionsout{\tr@ce}\expandafter\setfl@gs\tr@ce,}
\def\setfl@gs#1,{\def\@{#1}\ifx\@\empty\let\next=\relax
   \else\let\next=\setfl@gs\expandafter\xdef
   \csname#1tr@cetrue\endcsname{}\fi\next}
\newcount\sectno\sectno=0\newcount\subsectno\subsectno=0\def\r@s@t{\relax}
\def\resetall{\global\advance\sectno by 1\subsectno=0
  \gdef\firstpart{\number\sectno}\r@s@t}
\def\resetsub{\global\advance\subsectno by 1
   \gdef\firstpart{\number\sectno.\number\subsectno}\r@s@t}
\def\v@idline{\par}\def\firstpart{\number\sectno}
\def\l@c@l#1X{\firstpart.#1}\def\gl@b@l#1X{#1}\def\t@d@l#1X{{}}
\def\m@ketag#1#2{\expandafter\n@wcount\csname#2tagno\endcsname
     \csname#2tagno\endcsname=0\let\tail=\alltgs\xdef\alltgs{\tail#2,}%
  \ifx#1\l@c@l\let\tail=\r@s@t\xdef\r@s@t{\csname#2tagno\endcsname=0\tail}\fi
   \expandafter\gdef\csname#2cite\endcsname##1{\expandafter
     \ifx\csname#2tag##1\endcsname\relax?\else{\rm\csname#2tag##1\endcsname}\fi
    \expandafter\ifx\csname#2tr@cetrue\endcsname\relax\else
     \write\cit@tionsout{#2tag ##1 cited on page \folio.}\fi}%
   \expandafter\gdef\csname#2page\endcsname##1{\expandafter
     \ifx\csname#2page##1\endcsname\relax?\else\csname#2page##1\endcsname\fi
     \expandafter\ifx\csname#2tr@cetrue\endcsname\relax\else
     \write\cit@tionsout{#2tag ##1 cited on page \folio.}\fi}%
   \expandafter\gdef\csname#2tag\endcsname##1{\global\advance
     \csname#2tagno\endcsname by 1%
   \expandafter\ifx\csname#2check##1\endcsname\relax\else%
\fi
   \expandafter\xdef\csname#2check##1\endcsname{}%
   \expandafter\xdef\csname#2tag##1\endcsname
     {#1\number\csname#2tagno\endcsnameX}%
   \write\t@gsout{#2tag ##1 assigned number \csname#2tag##1\endcsname\space
      on page \number\count0.}%
   \csname#2tag##1\endcsname}}%
\def\m@kecs #1tag #2 assigned number #3 on page #4.%
   {\expandafter\gdef\csname#1tag#2\endcsname{#3}
   \expandafter\gdef\csname#1page#2\endcsname{#4}}
\def\re@der{\ifeof\t@gsin\let\next=\relax\else
    \read\t@gsin to\t@gline\ifx\t@gline\v@idline\else
    \expandafter\m@kecs \t@gline\fi\let \next=\re@der\fi\next}
\def\t@gs#1{\def\alltgs{}\m@ketag#1e\m@ketag#1s\m@ketag\t@d@l p
    \m@ketag\gl@b@l r \n@wread\t@gsin\openin\t@gsin=\jobname.tgs \re@der
    \closein\t@gsin\n@wwrite\t@gsout\openout\t@gsout=\jobname.tgs }
\outer\def\localtags{\t@gs\l@c@l}
\outer\def\globaltags{\t@gs\gl@b@l}
\outer\def\newlocaltag#1{\m@ketag\l@c@l{#1}}
\outer\def\newglobaltag#1{\m@ketag\gl@b@l{#1}}

\def\t@gsoff#1,{\def\@{#1}\ifx\@\empty\let\next=\relax\else\let\next=\t@gsoff
   \expandafter\gdef\csname#1cite\endcsname{\relax}
   \expandafter\gdef\csname#1page\endcsname##1{?}
   \expandafter\gdef\csname#1tag\endcsname{\relax}\fi\next}
\def\verbatimtags{\let\ift@gs=\iffalse\ifx\alltgs\relax\else
   \expandafter\t@gsoff\alltgs,\fi}
\catcode`\X=11 \catcode`\@=\active
\localtags
%

%

%
\centerline{\bf Kinetics of a Model Weakly Ionized Plasma}
\centerline{\bf in the Presence of Multiple Equilibria}
\vskip .3cm
\centerline {by}
\centerline{E. Carlen\footnote{$^1$}{School of Mathematics, Georgia
Institute of Technology, Atlanta, GA 30332--0160}, R.
Esposito\footnote{$^2$}{Dipartimento di Matematica, Universit\`a di 
L'Aquila, Coppito---67100 L'Aquila---Italy}, J.L.
Lebowitz\footnote{$^3$}{Department of Mathematics and Physics, Rutgers
University,  New Brunswick, NJ 08903}, R.
Marra\footnote{$^4$}{Dipartimento di Fisica, Universita' di Roma Tor
Vergata, Via della Ricerca Scientifica, 00133 Roma, Italy} and A.
Rokhlenko{$^3$}}
\bigskip

\bigskip
\bigskip {\baselineskip = 12pt\narrower{\noindent {\bf Abstract:}\/} 
We study, globaly in time, the velocity distribution $f(v,t)$ of a
spatially homogeneous system that models a system of electrons in a
weakly ionized plasma, subjected to a constant external electric field
$E$. The density $f$ satisfies a Boltzmann type kinetic equation
containing a full nonlinear electron-electron collision term as well
as linear terms representing collisions with reservoir particles
having a specified Maxwellian distribution. We show that when the
constant in front of the nonlinear collision kernel, thought of as a
scaling parameter, is sufficiently strong, then the $L^1$ distance
between $f$ and a certain time dependent Maxwellian stays small
uniformly in $t$.  Moreover, the mean and variance of this time
dependent Maxwellian satisfy a coupled set of nonlinear ODE's that
constitute the ``hydrodynamical'' equations for this kinetic system.
This remain true even when these ODE's have non-unique equilibria,
thus proving the existence of multiple stabe stationary solutions for
the full kinetic model. Our approach relies on scale independent
estimates for the kinetic equation, and entropy production
estimates. The novel aspects of this approach may be useful in other
problems concerning the relation between the kinetic and hydrodynamic
scales globably in time.

\bigskip
\bigskip
\bigskip} 

\noindent{\bf 1. Introduction}
\numsec= 1
\numfor= 1 
\vskip.2cm

The mathematical understanding of equilibrium phenomena has greatly
advanced in the past few decades.  The elegant and precise theory of
Gibbs measures provides a direct bridge between the microscopic and
macroscopic descriptions of such systems.  This includes a general
conceptual framework as well as nontrivial explicit examples of the
coexistence of multiple equilibrium phases for certain values of the
macroscopic control parameters.

There is no comparable general theory for nonequilibrium systems and
the microscopic study of phase transition phenomena in such situations
appears to be far beyond our mathematical grasp at the present time.
Our mathematical understanding of the great variety of nonequilibrium
phase transitions observed in fluids, plasmas, lasers, etc., therefore
depends entirely on the study of bifurcations and other singular
phenomena occurring in the nonlinear equations describing the {\it
macroscopic} time evolution of such systems.

While there has been much progress recently in deriving such equations
from simple microscopic and even realistic mesoscopic model evolutions
the passage to the macroscopic scale is well understood {\it only over
time intervals in which the solutions of the macroscopic equations
stay smooth}. This is true for example in the passage from kinetic
theory, where the evolution is described by the Boltzmann equation, to
hydrodynamics, where it is described by either the compressible Euler
or Navier-Stokes equations, depending on how we choose our macroscopic
time scale [\rcite{N}, \rcite{Ca}, \rcite{U}, \rcite{DEL},
\rcite{ELM1}, \rcite{ELM2}].  These derivations, which are based on
Chapman-Enskog type expansions, require for their validity the
uniqueness and smoothness of the solutions of the hydrodynamic
equations.  The reason for this is that control of the error terms in
the estimates depends on {\it a-priori} smoothness estimates for
solutions of the macroscopic equations.  Thus, they shed no light on
the actual behavior of the mesoscopic description when the solution of
the hydrodynamical equations develop singular behavior.

To overcome this problem it is clearly desirable to develop methods in
which one does not use any {\it a-priori} smoothness estimates for
solutions of the macroscopic equations, but instead uses scale
independent estimates on the mesoscopic equation. This is what we do
here for a simple model inspired by plasma physics [\rcite{R},
\rcite{B}, \rcite{F}].

Our starting point is a kinetic theory description of the
system. Grave difficulties are posed by the fact that as of yet, not
very much is known in the way of {\it a-priori} regularity estimates
for solutions of the spatially inhomogeneous Boltzmann equation. This
is quite different however, from the lack of estimates for the
macroscopic equations -- there it is clear that in the interesting
cases the desired estimates just don't exist. Shock waves do form. In
the Boltzmann case however, it is likely that {\it a-priori}
regularity estimates in the velocity variables, say, invariant under
the Euler scaling, are there, but simply have not yet been
discovered. Still, the lack of such estimates is a grave difficulty in
the way of rigorous investigation of the problem at hand.

We sidestep this difficulty by considering a spatially homogeneous
system, but one that is driven by an electric field, and coupled to
heat reservoirs.  In this case the usual hydrodynamic moments are not
conserved and the system will have non-equilibrium stationary
states. We prove then in a certain simplified, but still recognizable
physical situations, that the kinetic description closely tracks the
macroscopic description even when the driving is sufficiently strong
for the latter to undergo phase transitions.  More precisely, we show
that the velocity distribution function is close to a Maxwellian
parametrized by a temperature $T$ and mean velocity $u$ which satisfy
certain non-linear equations, which are the macroscopic equations for
this system. Moreover, it does so globally in time, even when the
stationary solutions of these macroscopic equations are nonunique.

We are in fact particularly concerned with the stability of of these
stationary solutions -- the existence of multiple stationary states
being analogous to the coexistence of phases in equilibrium systems.
{}For such questions we need results that guarantee that a solution of
the kinetic equations will stay near a solution of the macroscopic
equations globaly in time.  This seems to be difficult to accomplish
by standard expansion methods, at least in the range of driving field
strengths where the macroscopic equations have the most interesting
behavior.  Instead of expansion methods, we use entropy production
[\rcite{CC1}, \rcite{CC2}] to show that the solution of the kinetic
equations must stay close to {\it some} Maxwellian, globally in
time. Then, we show that the moments of this Maxwellian must nearly
satisfy the macroscopic equations. In this way we get our results. The
next section specifies the model more closely, and states our main
results. A preliminary account of this work in which the Boltzmann
collisions were modeled by a BGK collision kernel was presented in
[\rcite{CELMR}].

\bigskip

\noindent{\bf 2. The model and the results}
\bigskip
\numsec= 2
\numfor= 1 
\vskip.2cm 

Our formal set up is as follows: We consider a weakly ionized gas in
${\IR^3}$ in the presence of an externally imposed constant electric
field $E$. The density of the gas, the degree of ionization and the
strength of the field are assumed to be such that: (i) the
interactions between the electrons can be described by some nonlinear,
Boltzmann type collision operator, and (ii) collisions between the
electrons and the heavy components of the plasma, ions and neutrals,
are adequately described by assuming the latter ones to have a
spatially homogeneous time independent Maxwellian distribution with an
a proiri given temperature [\rcite{R}].  Under these conditions the
time evolution of the spatially homogeneous velocity distribution
function $f(v,t)$ will satisfy a Boltzmann type equation
$$
{\partial f(v,t) \over \partial t} = -E \cdot \nabla f + Lf +
\epsilon^{-1} Q(f),\Eq(2.1)$$
where $\nabla$ is the gradient with respect to $v$ in ${\IR^3}$, $E$
is a constant force field and $Q$ is a nonlinear collision term which
will take either the form of the Boltzmann collision kernel for
Maxwellian molecules, or the one corresponding to the BGK model. We
treat both cases here because it is possible to provide a little more
detail concerning the nature of the equilibria in the BGK case.  The
parameter $\epsilon> 0$ is thought of as a scaling parameter that goes
to zero in the hydrodynamical limit. The linear operator $L$
represents the effect of collisions with reservoir particles.  It will
be assumed to have the form:
$$Lf(v) = L_1f(v) +L_2f(v),\Eq(2.1.3)$$
with
$$L_1f(v)=\nabla\cdot\biggl(D(v)M(v)\nabla\biggl({f(v)\over M(v)}
\biggr)\biggr),\Eq(2.2)$$
a {}Fokker-Planck operator, representing energy exchanges with the
reservoir assumed to be at temperature $T = 1$, so that 
$$M(v) = (2\pi )^{-3/2}\exp(-|v|^2/2)$$
and
$$D(v) = a\exp(-b|v|^2/2) + c\Eq(2.3)$$ for some strictly positive
constants $a$, $b$ and $c$: the symbols $a$, $b$, $c$ shall henceforth
always refer to these parameters wherever they appear.  The specific
form \equ(2.3) of the velocity space diffusion coefficient is not
important. We specify it for sake of concreteness.  The properties we
really need for $D(v)$ will be clear from the proofs.  The operator
$L_2$ represents momentum exchanges with the heavy reservoir particles
and is given by
$$L_2f(v)= \nu[\bar f(v) -f(v)],\Eq(2.3.3)$$
with $\nu$ a positive constant and $\bar f(v)$ the sphericalized 
average of $f(v)$.

{}For any probability density $f$, we shall let $M_f$  denote the 
Maxwellian density with the same first and second moments as $f$.
Explicitly,
$$M_f= (2\pi T)^{-3/2} \exp[- {(v-u)^2\over 2T}],\Eq(2.3.5)$$
with
$$u := \int_{\IR^3}vf(v)\d,\Eq(2.4)$$
$$e := {1 \over 2}\int_{\IR^3}v^2f(v)\d ,\Eq(2.5)$$
and $T = {2 \over 3} (e - {1 \over 2} u^2)$.
In the BGK model the collision kernel is
$$Q_{BGK}(f) = M_f - f\ .$$
The Boltzmann collision kernel is given by 
$$Q_B(f)(v)= \int_{\IR^3}d\/v_*\int_{S_2^+}d\/\o
B(|v-v_*|,\o)\big[f(v')f(v'_*)-f(v)f(v_*)\big].\Eq(2.3.6)$$
Here $S_2^+=\{\o\in \IR^3: \o^2=1, \o\cdot (v-v_*)\ge 0\}$, and 
$$\eqalign
{
v'= &v-\o\cdot(v-v_*)\o,\cr
v'_*=&v_*+\o\cdot (v-v_*)\o,
}\Eq(2.3.61)
$$
are the outgoing velocities in a collision with incoming velocities
$v$ and $v_*$ and impact parameter $\o$, $B(|v-v_*|,\o)$ is the
collision cross section, depending on the intermolecular
interactions. {}For Maxwellian molecules, with a Grad angular cut-off
[Gr],
$$B(|v-v_*|,\o)= h(\theta),\Eq(2.3.7)$$
with $\theta$ the azimuthal angle of the spherical coordinates in $S_2$ 
with polar axis along $v-v_*$ and $h(\theta)$ a smooth non negative
bounded function.  Thus, for any normalized $f$ we can write
$$Q_B(f)=\ell\bigl(f\circ f - \/\/f\bigr)\Eq(2.3.8)$$
with
$$\eqalign
{
f\circ f(v)=&{1\over 2\ell}\int_{\IR^3}d\/v_*\int_{0}^{2\pi}d\/
\varphi\int_0^\pi d\/\theta\/ h(\theta)\/ |\sin\/\theta| 
f(v')f(v'_*),\cr
\ell=& \pi \int_0^\pi d\/\theta\/ h(\theta)\/ |\sin \/\theta|>0,
}
\Eq(2.3.9)$$
With either $Q_{BGK}$ or $Q_B$ for the collision kernel in \equ(2.1),
this term tends to keep $f$ close to $M_f$, and one could certainly
expect this effect to dominate for small values of $\epsilon$. Thus,
formally in the limit as $\epsilon$ vanishes, $f$ will actually equal
$M_f$ for all time $t$, and to keep track of its evolution, we need
only keep track of $u(t)$ and $e(t)$.

Using the prescription $f= M_f$, in the right side of (2.1), one
easily evaluates the time derivative of the first two moments of the
so modified (2.1), to obtain formally
$${{\rm d}\over {\rm d}t}\left(\matrix{\tilde u(t)\cr
\tilde e(t)\cr}\right) = \left(\matrix{F(\tilde u(t), \tilde e(t))\cr
G(\tilde u(t), \tilde e(t))\cr}\right),\Eq(2.5.3)$$
The functions $F$ and $G$ are given explicitly by 
$$
{}F(u,e) = E - u[\nu + c + a \exp(-w) {1+b \over (1+b/\beta)^{5/2}}], 
\Eq(2.5.5)
$$
$$
G(u,e) = Eu -c\Big[2e(1-\beta)+\beta u^2\Big] - {a \exp(-w) \over
(1 + b/\beta)^{{5 \over 2}}} \Big[2e(1-\beta) + u^2(\beta - 
b {1-\beta \over b+\beta})\Big], \Eq(2.5.7)
$$
where $\beta=T^{-1}$ and $w = bu^2/2(1+bT)$.  The tildes in (2.14) 
are to remind us that the equation is valid only when $f = M_f$.

Equations \equ(2.5.3) represent the hydrodynamical description of the
gas. Our primary goal here is to show that such a description actually
does hold for small, but positive, values of $\epsilon$, i.e. that the
interaction between the hydrodynamic and the non--hydrodynamic modes
does not destroy the picture involving only the hydrodynamic
modes. The following theorem enables us to do this.
\bigskip
 
\noindent{\bf Theorem 2.1.}  {\it Let $f$ be a solution of \equ(2.1) 
with 
$$f(\cdot,0) = M_{f(\cdot,0)}\ .$$
Then, for any fixed integer $k_0>0$, there is an $\e_0>0$ and  
functions
$\delta_1(\e)$, $\delta_2(\e)$, going to zero  as $\e\to 0$, depending 
only on
$a$, $b$, $c$, $|E|$ and $e(0)$, such that for $\e<\e_0$ the solution 
of
\equ(2.1) satisfies
$$\sup_{t\in \IR^+}\|f(\cdot,t) - M_{f(\cdot,t)}\|_{L^1(\IR^3)} \le
\delta_1(\e)\Eq((2.7)$$ and
$$\sup_{t\in \IR^+}\|\nabla^k\bigl(f(\cdot,t) -
M_{f(\cdot,t)}\bigr)\|_{L^2(\IR^3)} 
\le\delta_2(\e)\Eq((2.8)$$}
for $k\le k_0$.
\bigskip
 
\noindent{\bf Remark:}  The assumption on $f(\cdot,0)$ is not 
essential: our methods allow us to easily modify the result to take
into account an initial layer.
\bigskip
 
We shall use Theorem 2.1, in the proof of Theorem 2.2 below, that if
we compute $u(t)$ and $e(t)$ for a solution $f$ of \equ(2.1)
satisfying the conditions of Theorem 2.1, then the moments of $f$ will
satisfy the equation,
$${{\rm d}\over {\rm d}t}\left(\matrix{u(t)\cr
e(t)\cr}\right) = \left(\matrix{F(u(t), e(t))\cr
G(u(t), e(t))\cr}\right) + \delta(\e)^{1/2}
\left(\matrix{\gamma(t)\cr
\eta(t)\cr}\right)\Eq(2.9)$$
where $F$ and $G$ are the non--linear functions of $u$ and  $e$ given 
by
\equ(2.5.5) and \equ(2.5.7), that arise in the $\e = 0$ limit, and
$\gamma(t)$ and
$\eta(t)$ are bounded uniformly in $t$ with a bound independent of 
$\e$. Of course they depend on the full solution $f$ of \equ(2.1), but
such estimates and a simple  comparison argument will then lead to the
following theorem, which says that the system (2.14) obtained in the 
$\epsilon \to 0$ limit does give an
accurate picture of the small $\epsilon$ regime.
\bigskip 
 
\noindent{\bf Theorem 2.2 } {\it Let $(u^*,e^*)$ be a stable fixed
point of the system \equ(2.5.3) and let $M_{(u^*,\e^*)}$ be the
corresponding Maxwellian density, with moments $u^*$ and $e^*$. Then
given any $\delta>0$, there is an $\epsilon$ greater than zero such
that if $f_\e(\cdot,t)$ solves \equ(2.1) with this value of $\e$ and
$$\|f_\epsilon(\cdot,0) - M_{(u^*,e^*)}(\cdot)\|_{L^1(\IR^3)} \le
\epsilon\Eq(2.10)$$ 
then
$$\|f_\epsilon(\cdot,t) - M_{(u^*,e^*)}(\cdot)\|_{L^1(\IR^3)} \le
\delta\Eq(2.11)$$ 
for all $t\ge 0$.
 
If, however, $(u^*,e^*)$ is not stable, then there exist a $\delta >0$
so that for every $\epsilon >0$, there is a solution
$f_\epsilon(\cdot,t)$ of \equ(2.1) with Maxwellian initial data
satisfying
$$\|f_\epsilon(\cdot,0) - M_{(u^*,e^*)}(\cdot)\|_{L^1(\IR^3)} \le
\epsilon\ ,\Eq(2.12)$$ 
but such that for some finite $t>0$
$$\|f_\epsilon(\cdot,t) - M_{(u^*,e^*)}(\cdot)\|_{L^1(\IR^3)} \ge 
\delta\ .\Eq(2.13)$$}
\bigskip

The proof of these theorems, which are fairly complicated even for 
the simple  BGK model will be given in the next sections.

The above theorems allow us to rigorously prove that our kinetic
system has multiple equilibria in certain ranges of the parameters
that specify it.  This is because of the following result concerning
the ``hydrodynamic'' system (2.14).
\bigskip

\noindent{\bf Proposition 2.3 } 
{\it There are choices of the parameters $a$, $b$, $c$ and $\nu$ for
which there are nonempty intervals $(E_0,E_1)$ such that, if $|E|$ is
outside of the closed interval $[E_0,E_1]$, then there is unique
stable fixed point for the system \equ(2.5.3), while, if $|E|\in
(E_0,E_1)$ then there are three fixed points for the system
\equ(2.5.3). Moreover two of them are stable and one is unstable. 
}\bigskip

Stability here is meant in the sense that the eigenvalues of the
differential have a strictly negative real part.
 
The proof of Proposition 2.3 is an explicit calculation which we omit
(see [7]). However, to gain an intuitive understanding of why there
should be multiple stable equilibria for certain parameter ranges,
think of the $\e=0$ limit of \equ(2.1) as a constrained motion on the
``manifold of Maxwellians''. Without this constraint, which is
generated by the collision kernel, the evolution would be the one
ruled by the electric field and the linear operator $L$. Clearly, this
evolution has an unique attracting fixed point, which, for $a>0$, is
not Maxwellian.  What happens is that there are one or more places on
the {\it non-linear} constraint manifold that are {\it locally
closest} to the attracting point of the unconstrained system. Each of
these is a stable equilibrium for the constrained evolution. As the
parameters are varied, the position of the unconstrained fixed point
relative to the manifold of Maxwellians varies, and with this
variation in geometry, the number of locally closest points varies.

The main physical issues regarding this model are sttled at this
point: We have proved the existence of the multiple stable equilibria
at the kinetic level -- $\epsilon$ small, but positve -- that had been
found and investigated in [7] at $\epsilon = 0$. Moreover, we remind
the reader that we do not know how to establish such a result using
conventional expansion methods: the difficulty being that if $E$ is
not small, and hence possibly out of the range where multiple
equilibria exist for (2.14), we only know how to prove (2.21) {\it
locally} in time. This is insufficient to show that for $\epsilon$
small enough, one {\it never} wanders far from any $M_{(u^*,e^*)}$
with $(u^*,e^*)$ stable for (2.14). While the entropy methods we use
do let us do this for arbitrary $E$, there are several finer question
that one could ask, but for which we have only incomplete answers.

{}First, one can ask whether or not there is an actual stationary
solution inside the invariant neighborhoods that we have found of the
$M_{(u^*,e^*)}$, and second, once one knows that stationary solutions
exist, one can ask whether or not solutions actually tend to converge
to one of these stationary solutions as $t$ tends to infinity.

The first question we can answer comletely only in the BGK case. The positive
answer is given by:
 
\medskip

\noindent{\bf Theorem 2.4} {\it

Let $Q=Q_{BGK}$ and $a>0$. Then for each fixed point of \equ(2.5.3)
there is exactly one stationary solution of eq. \equ(2.1).  This
solution lies in a suitably smalll neighborhood of $M_{u^*,e^*}$, the
Maxwellian corresponding to $(u^*,e^*)$, and it inherits the stability
properties of the hydrodynamical fixed point.  }
\medskip

{}For the Boltzmann kernel we have only a partial result that is
reported in Section 8. Also on the question of convergence we have
only very partial results. These are reported in Section 9, where the
difficulties are explained as well.  But though it would be desirable
to have a more complete resolution of these issues, they are not
central to establishing that the kinetic systems does actually have
the several stable regimes that one sees in the limiting
``hydrodynamic'' equations.
 
The proofs are organized as follows. In Section 3 we prove moment
bounds. Section 4 contains the proof of two ``interpolation
inequalities''. The first of these will be used to obtain {\it
a--priori} smoothness bounds in Section 5. The second will be used to
transform the smoothness bounds of Section 5 into a lower bound on the
variance of our density. The smoothness requires Sobolev estimates for
the collision kernel, which are straightforward for $Q_{BGK}$, while
for $Q_B$ they rely on some recent results [\rcite{CGT}], which we
simply state here.  Having assembled these moment, smoothness and
interpolation bounds, we can use a key entropy production inequality
for $Q_B$ proved in [\rcite{CC2}]. The analogous inequality for
$Q_{BGK}$ is proven in a simpler way in Section 6. This is used to get
quantitative bounds on the tendency of the collision operator to keep
the density nearly Maxwellian. What we obtain directly is $L^1$
control on the difference between $f$ and $M_f$, but the smoothness
bounds together with the interpolation bounds allows us to obtain
control in stronger norms. Section 7 contains the proofs of the
Theorems 2.1 and 2.2, which, given the lemmas, are quite short.
Section 8 is devoted to the proof of existence of stationary
solutions, In section 9 we discuss the tendancy toward these
stationary solutions.
 
\bigskip
\noindent{\bf 3. Moment Bounds}
\bigskip
\numsec= 3
\numfor= 1 
\vskip.2cm 

In this section we establish {\it a--priori} moment bounds for
solutions of (2.1). In estimating the evolution of the moments, we
shall use one set of methods to treat the effects of collisions, and
another set to treat everything else. Thus it is natural to rewrite
(2.1) as
$${\partial\over\partial t}f(v,t) = {\cal L}f(v,t) + {1\over \epsilon}
Q(F)(v,t)\Eq(3.1)$$ 
where  ${\cal L}f = -E\cdot\nabla f 
+ Lf$.

We shall use the standard ``bracket notation'' notation for averages:
$\langle \phi \rangle_t$ denotes $\int_{\IR^3}\phi(v)f(v,t){\rm d}^3v$
for any positive or integrable function $\phi$.

Throughout this paper, $K$ will denote a computable constant that
depends at most on the electric field $E$, the paramters $a$, $b$, $c$
and $\nu$ specified in \equ(2.3), and where indicated, also on the
fourth moment of the initial distribution: $\langle |v|^4 \rangle_0$.
The constant will, however, change from line to line.

In these terms, the main result of this section is:
\bigskip

\noindent{\bf Theorem 3.1}\  {\it Let $f$ denote a solution to (2.1). 
Then for
$Q=Q_B$ or $Q=Q_{BGK}$, we have
$$\langle |v|^4 \rangle_t \le \langle |v|^4 \rangle_0 + K\ .$$}
\bigskip

Note that by Jensen's inequality, this imediately controls all lower 
order moments as well. The first step, however,  is to directly conrol
the second moments.  
 
\bigskip
\noindent{\bf Lemma 3.2}\ {\it Let $f$ denote a solution to \equ(2.1). 
Then for both choices of the collision kernel $Q$, we have
$$\langle |v|^2 \rangle_t \le \langle |v|^2 \rangle_0 + K\Eq(3.2)$$}  

\bigskip
 
\noindent{\bf Proof:} Since $\langle |v|^2 \rangle_t$ is a collision 
invariant, and $\int  v^2 L_2f{\rm d}^3v=0$,  
$$\eqalign{&{{\rm d}\over {\rm d}t}\langle |v|^2 \rangle_t = 
\int_{\IR^3}|v|^2{\cal L}f(v,t)\d\cr &+2E\cdot\langle v \rangle_t  -2
\int_{\IR^3}D(v)M(v)v\cdot\nabla\biggl({f(v,t)\over M(v)}\biggr) {\rm
d}^3v =\cr &2E\cdot\langle v \rangle_t + 6\langle D \rangle_t
+2(v\cdot\nabla D)\langle (v\cdot\nabla D) \rangle_t 
-2\langle D|v|^2 \rangle_t\cr}$$
 
Now observe that 
$$(v\cdot\nabla D) \le 0\Eq(3.3)$$ 
 and that
$$c \le D \le (a+c)\Eq(3.4)$$
for all $v$. {}Finally, by Jensen's inequality, 
$$|\langle v \rangle_t| \le
\bigl(\langle |v|^2 \rangle_t\bigr)^{1/2}\Eq(3.5)$$
These facts, combined with the previous calculation, yield the 
estimate
$${{\rm d}\over {\rm d}t}\langle |v|^2 \rangle_t \le 
-2c\langle |v|^2 \rangle_t +6(a+c) +2|E|\bigl(\langle |v|^2 \rangle_t
\bigr)^{1/2}$$  Straightforward estimation now leads to
$${{\rm d}\over {\rm d}t}\langle |v|^2 \rangle_t \le 
-c\langle |v|^2 \rangle_t  + K\ .$$
Then \equ(3.2) in turn follows from the fact that any solution of the
differential inequality
$\dot x(t) \le -cx(t) + K$ satisfies $x(t) \le x(0) + K/c$. \ $\square$
\bigskip
 
We next parley these bounds into bounds on the fourth moments; i.e., 
$\vf$. Since 
$\vf$ is not a collision invariant, these depend on the particular 
collision kernel under consideration.

\bigskip
\noindent{\bf Lemma 3.3}\ {\it Let $Q=Q_{BGK}$. Then
for any density $f$,
$$\int_{\IR^3}|v|^4Q(f)\d \le (160/3)\bigl(\vt\bigr)^2 - \vf$$}
\bigskip 
\noindent{\bf Proof:} By an easy calculation, 
$$\int_{\IR^3}|v -\langle v \rangle_t|^4M_f(v)\d  =$$
$$ (5/3)\biggl(\int_{\IR^3}|v -\langle v \rangle_t|^2M_f(v)\d\biggr)^2 
=  (5/3)\langle|v -\langle v\rangle_t|^2 \rangle^2 _t$$
Next, note that
$|v| \le |v-\langle v \rangle_t| + |\langle v \rangle_t|$, and thus,
$|v|^4 \le 16(|v-\langle v \rangle_t|^4 + |\langle v \rangle_t|)^4$.
Combining this and Jensen's inequality as in \equ(3.5) with the above,      
we have
$$\int_{\IR^3}|v|^4M_f(v)\d \le (160/3)\bigl(\vt\bigr)^2\ .$$
The result now follows directly from the form of $Q(f)$.\quad 
$\square$
\bigskip
\noindent{\bf Lemma 3.4}\ {\it Let $Q=Q_{B}$. Then there are positive
constants $c_1$ and $c_2$ such that for any density $f$,
$$\int_{\IR^3}|v|^4Q(f)\d \le c_1\bigl(\vt\bigr)^2 - c_2\vf$$}
\bigskip 
\noindent{\bf Proof:} The result follows once we prove that 
$$\int d^3 v |v|^4 f\circ f(v) \le  (\ell - c_2)\vf + 
c_1\bigl(\vt\bigr)^2.$$ To check this, let $\pi_\o$ and $\pi^\perp_\o$
denote the projection in the direction $\o$ and the complementary
projection respectively. Then we can write $v$ as
$$v= \pi_\o v_* + \pi^\perp_o v'_*,$$ 
and
$$|v|^2= |\pi_\o v_*|^2 + |\pi^\perp_o v'_*|^2.$$
Hence 
$$|v|^4 = |\pi_\o v_*|^4 + |\pi^\perp_o v'_*|^4 + 2|\pi_\o v_*|^2 
|\pi^\perp_o v'_*|^2.$$
Averaging on $S_2$ we get
$$\eqalign
{
&\int_{s_2} d\/\o B(|v-v_*|,\o)[|\pi_\o v_*|^4 +|\pi^\perp_o v'_*|^4
]  = |v|^4 \pi\int_0^\pi d\/\theta |\sin\/\theta |\/h(\theta) 
[\cos^4 \theta +\sin^4 \theta]\cr
&= |v|^4\Big( \ell - 2 \pi\int_0^\pi d\/\theta |\sin\/\theta 
|\/h(\theta) 
\cos^2 \theta \sin^2 \theta\Big)\equiv |v|^4( \ell -c_2 ).
}
$$
\ $\square$
\bigskip 
 
\noindent{\bf Proof of Theorem 3.1:} Calculating as before, we have:

$$\eqalign{&{{\rm d}\over {\rm d}t}\langle |v|^4 \rangle_t = 
\int_{\IR^3}|v|^4{\cal L}f(v,t)\d + {1\over \e}
\int_{\IR^3}|v|^4Q(f)(v,t)\d=\cr &4E\cdot\langle |v|^2v \rangle_t  
-4\int_{\IR^3}D(v)M(v)|v|^2v\cdot\nabla\biggl({f(v,t)\over M(v)}\biggr)
{\rm d}^3v + {1\over \e}\int_{\IR^3}|v|^4Q(f)\d=\cr &4E\cdot\langle
|v|^2v \rangle_t + 12\langle |v|^2D \rangle_t +4(|v|^2v\cdot\nabla
D)\langle (v\cdot\nabla D) \rangle_t -2\langle D|v|^4 \rangle_t + 
{1\over\epsilon}\int_{\IR^3}|v|^4Q(f)\d\cr\ .}$$ 

We now estimate using \equ(3.3) and \equ(3.4) just as in the proof of
Lemma 3.2, together with Lemmas 3.3 and 3.4, to control the collision
term. (Note that once again $L_2$ makes no contribution.)  The result
is
$$\eqalign{&{{\rm d}\over {\rm d}t}\langle |v|^4 \rangle_t \le\cr
&4|E|\bigl(\vf\bigr)^{3/4} + 12(a+c)\vt - 4c\vf +{1\over \epsilon}
\bigl(c_1\bigl(\vt\bigr)^2 - c_2\vf\bigr)\ .\cr}$$ Next, by Lemma 3.1 
together with Jensen's inequality,  we know that $12(a+c)\vt$ is
bounded above by a universal constant plus $\bigl(\langle
|v|^4\rangle_0\bigr)^{1/2}$. Thus, if we introduce the  scaled time
parameter
$$\tau = (1/\epsilon)t\ ,$$
and define $x(\tau) := \langle |v|^4\rangle_\tau$, we have that
$x(\tau)$ satisfies a differential inequality of the form
$$\dot x(\tau) \le -x(\tau) + K$$
from which the result follows.\quad$\square$


\bigskip
\noindent{\bf 4. Interpolation inequalities}
\bigskip
\numsec= 4
\numfor= 1 
\bigskip
 
The lemmas in this section are several interpolation inequalities
related to the familiar Gagliardo--Nirenberg inequalites, but with
some special features adapted to our applications. The inequality of
Lemma 4.2 is the most novel and interesting of these.
 
\bigskip
 
\noindent{\bf Lemma 4.1}\ {\it Let $f\in L^1(\IR^3)$. Then there is a
universal constant $C$ such that if $f$ has a square integrable
distributional Laplacean, then $f$ has a square integrable gradient,
and
$$\|\nabla f\|_2 \le C\|f\|_1^{2/7}\|\Delta f\|_2^{5/7}\Eq(4.1)$$
Similarly, there is a universal constant $C$ such that if $\Delta^2f$
is square integrable, then $\Delta f$ is also square integrable, and
 $$\|\Delta f\|_2 \le C\|f\|_1^{4/11}\|\Delta^2 f\|_2^{7/11}\Eq(4.2)$$}
\bigskip
 
\noindent{\bf Proof:}\  Taking {}Fourier transforms, we have
$$\eqalign{&\|\nabla f\|_2^2 = \int_{\IR^3}|p|^2|\hat f(p)|^2{\rm d}^3p =\cr
&\int_{|p|\le R}|p|^2|\hat f(p)|^2{\rm d}^3p + 
\int_{|p|\ge R}|p|^2|\hat f(p)|^2{\rm d}^3p\le \cr
&CR^5\|f\|_1^2 + R^{-2}\int_{|p|\ge R}|p|^2|\hat f(p)|^4{\rm d}^3p
\le\cr &CR^5\|f\|^2_1 + R^{-2}\|\Delta f\|_2^2\cr}$$
where the computable constant $C$ changes from line to line.
Optimizing over $R$ now yields \equ(4.1). The proof of \equ(4.2) is
done in the same way. \quad $\square$
 
\bigskip
 
The next inequality is similar in effect to an ``uncertainty principle''.
We shall use it to obtain uniform lower bounds on the variance of our
density $f$.
 
\bigskip
 
\noindent{\bf Lemma 4.2}\ {\it Let $f\in L^1(\IR^3)$, and suppose that
$f$ has a square integrable distributional gradient. Then there is a 
universal constant $C$ such that
$$\int_{\IR^3}f(v)\d \le C\|\nabla f\|_2^{4/9}\biggl(\int_{\IR^3}
\biggl|v - \int_{\IR^3}vf(v)\d\biggr|^2f(v)\d\biggr)^{5/9}\Eq(4.3)$$}

\bigskip
 
\noindent{\bf Proof:}\ The right side of \equ(4.3) is decreased when 
we replace $f$ by its spherically symmetric decreasing rearrangement,
while the left side is unchanged. We may therefore assume without loss
of generality that $f$ is spherically symmetric and radialy
decreasing. Now fix $R>0$, and define $g(v) := f(v)$ for $|v|\le R$,
and $g(v) = 0$ otherwise. Define $h$ by $f= g + h$. Now clearly,
$$\int_{\IR^3}h(v)\d \le R^{-2}\int_{|v|>R}|v|^2f(v)\d\ .$$
Also,
$$\eqalign{&\int_{\IR^3}g(v)\d =\cr
&(4\pi/3)R^3f(R) + (4\pi/3)R^3\biggl((4\pi/3)^{-1}R^{-3}\int_{|v|<R}
\bigl(f(v)-f(R)\bigr)\d\biggr)\le\cr
&(4\pi/3)R^3f(R) + (4\pi/3)R^3\biggl((4\pi/3)^{-1}R^{-3}\int_{|v|<R}
\bigl(f(v)-f(R)\bigr)^2\d\biggr)^{1/2}\cr}$$
Now, since $f$ is monotone, 
$$\int_{|v|<R}|v|^2f(v)\d \ge f(R)\int_{|v|<R}|v|^2f\d = 
f(R)(4\pi/5)R^5\ .$$ Then, with $\lambda$ denoting the principle
eigenvalue for the Dirichlet Laplacean in the unit ball, we have that
$$ \int_{|v|<R}\bigl(f(v)-f(R)\bigr)^2\d \le R^2\lambda^{-1}
\int_{|v|<R}\bigl|\nabla\bigl(f(v)-f(R)\bigr)\bigr|^2\d \le
R^2\lambda^{-1}\|\nabla f\|_2^2\ .$$
Combining the above, we have
$$\int_{\IR^3}f(v)\d \le C\biggl(\|\nabla f\|_2R^{5/2} + R^{-2}
\int_{\IR^3} |v|^2f(v)\d\biggr)$$
Optimizing over $R$ yields the result.\quad$\square$
\bigskip
 
\bigskip
\noindent{\bf 5. Smoothness bounds}
\bigskip
\numsec= 5
\numfor= 1 

The purpose of this section is to establish {\it a--priori} smoothness 
bounds for solutions of (2.1). The main result is the the following:

\noindent{\bf Theorem 5.1} {\it Let $f$ be a solution of \equ(2.1) such that 
$\|\nabla f(\cdot,0)\|_2$ is finite. Then, if $Q=Q_{BGK}$, there is a 
constant
$K$ such that
$$\|\nabla f(\cdot,t)\|_2 \le 
K\bigl( 1 + \|\nabla f(\cdot,0)\|_2\bigr)\Eq(5.1)$$
for all $t>0$. Similarly, suppose that 
$\|\Delta f(\cdot,0)\|_2$ is finite. Then there is a constant $K$ such 
that
$$\|\Delta f(\cdot,t)\|_2 \le 
K\bigl( 1 + \|\Delta f(\cdot,0)\|_2\bigr)\Eq(5.2)$$
for all $t>0$.

If $Q=Q_B$ the same results hold provided $\|f(\cdot,0) - 
M_{f(\cdot,0)}\|_1$ and $\epsilon$ are both sufficiently small.
}
 
\bigskip

{\bf Remark:} The smallness condition on $\|f(\cdot,0) -
M_{f(\cdot,0)}\|_1$ poses no problem here since we are avoiding an
initial layer by assuming that $f(\cdot,0) = M_{f(\cdot,0)}$. However,
it seems likely that it would be strightforward to include an inital
layer analysis, and to dispense with this condition -- even the
present proof does not require $\|f(\cdot,0) - M_{f(\cdot,0)}\|_1$ to
be particularly small.
 
\noindent{\bf Proof:}
Once again, we write (2.1) in the form (3.1). Then differentiating,
and integrating by parts, we have
$${{\rm d}\over {\rm d}t}\|\nabla f(\cdot,t)\|^2 = -2\int_{\IR^3}
\Delta f(v,t)
\biggl({\partial f(v,t)\over \partial t}\biggr)\d =$$
$$ -2\int_{\IR^3}\Delta f(v,t){\cal L}f(v,t)\d  -2\int_{\IR^3}\Delta 
f(v,t)Q(f)(v,t)\d\ .$$

As with the moment bounds, we begin by estimating the individual 
contributions to the ${\cal L}$ term.
\bigskip

\noindent{\bf Lemma 5.2}\ {\it There is a universal constant $K$  so
that for all solutions $f$ of (2.1),
$$ -2\int_{\IR^3}\Delta f(v,t){\cal L}f(v,t)\d \le K$$
for all $t \ge 0$}
\bigskip

{}First, $2E\cdot\int_{\IR^3}\Delta f\nabla f\d = 0$.
 
Next,  $$-2 \int_{\IR^3}\Delta f\bigl(\nabla D\cdot \nabla f)\d =
\int_{\IR^3}|\nabla f|^2(\Delta D|\d \le (a+c)\|\nabla f\|_2^2$$
 
Now, repeatedly integrating by parts:
$$\eqalign{&-2\int_{\IR^3}\Delta f\bigl(\nabla D\cdot v\bigr)f\d =\cr
&2\int_{\IR^3}\nabla f\bigl(v(\Delta D)f +3 \nabla D f + (v\cdot
\nabla D) \nabla 
f\bigr)\d \le \cr
&-3\int_{\IR^3}f^2(\Delta D)\d  -\int_{\IR^3}f^2\bigl(v\cdot
\nabla(\Delta D)
\bigr)\d\cr}$$
where we have once again used \equ(3.3). This gives us a bound of the 
form
$$-2\int_{\IR^3}\Delta f\bigl(\nabla D\cdot v\bigr)f\d \le K\|f\|_2^2\ .$$
We now use a standard interpolation inequality, the Nash inequality:
$$\|f\|_2 \le C\|\nabla f\|_2^{3/5}\|f\|_1^{2/5}\ .$$
This allows us to eliminate $\|f\|_2$ in favor of $\|\nabla f\|_2$,
the quantity of interest, and $\|f\|_1$, the conserved quantity.
$$-2\int_{\IR^3}\Delta f\bigl(\nabla D\cdot v\bigr)f\d \le K
\|\nabla f\|_2^{6/5}\|f\|_1^{4/5}\ .$$
Next:
$$-2\int_{\IR^3}\Delta f\bigl(Dv\cdot \nabla f\bigr)\d = 
\int_{\IR^3}|\nabla f|^2\nabla\cdot(Dv)\d \le K\|\nabla f\|_2^2$$
Apart from the collision term and the favorable dissipation term, the 
only  other term to be bounded is
$-2\int_{\IR^3}\Delta fDf$. However, integrating by parts, we obtain 
terms  identical to terms that we have already bounded.

The dissipation term is bounded using Lemma 4.1 as follows:
$$\eqalign{-&2\int_{\IR^3}D(\Delta f)^2\d \le -2c\|\Delta f\|_2^2\le\cr
&K\|\nabla f\|_2^{14/5}\|f\|_1^{-4/5} = K\|\nabla f\|_2^{14/5}\cr}$$
since $\|f\|_1 = 1$ for all times $t$.

Thus, with $x(\tau) := \|f(\cdot,\tau)\|_2^2$, we have established
$$-2\int_{\IR^3}\Delta f(v,t){\cal L}f(v,t)\d \le K(-x(t)^{7/5} + 
K\bigl(x(t)^{3/5} + x(t)^{3/10} + x(t)\bigr)\bigr)
\le K$$
since the largest power if $x(t)$ has a negative coefficient. \ 
$\square$

\bigskip

To conclude the proof of \equ(5.1), we need smoothness bounds for the
collision kernel; in particular, we need an estimate on the smoothness 
of the gain term in the collision kernel. 

{}For the BGK case, this is given by the following:
\bigskip

\noindent{\bf Lemma 5.3:}\ {\it {}For any positive integer $n$ there is a 
constant $K$ depending only on the second moment of $f$, such that
$$ 2\int_{\IR^3}(-\Delta)^{n} f(v,t)Q(f)(v,t)\d  \le K - \|(-\Delta)^{n/2}
f\|_2^2\Eq(5.3)$$}
\bigskip
\noindent {\bf Proof:} We only check the case $n=1$; the rest are similar. 
We have
$$\eqalign{-&2\int_{\IR^3}\Delta fQ_{BGK}(f)\d = \cr
-&2\int_{\IR^3}(\Delta M_f)f\d - 2\|\nabla f\|_2^2\ .
}\Eq(5.3.3)$$
Now, since 
$$-2\int_{\IR^3}(\Delta M_f)f\d\le C T_f^{-1}\|\n f\|_2,$$
we can use Lemma 4.2 to get the bound
$$ T_f^{-1}\le C \|\n f\|_2^{4/5}\Eq(5.3.5)$$
and hence 
$$-2\int_{\IR^3}(\Delta M_f)f\d\le C T_f^{-1}\|\n f\|_2\le C\|\n
f\|_2^{9/5}\le K+ \|\n f\|_2^2.$$
Combining this with \equ(5.3.3) we get the Lemma 5.3 for $n=1$\ 
$\square$
\bigskip

The analog of Lemma 5.3 for $Q_B$ is slightly more complicated.

\bigskip

\noindent{\bf Lemma 5.4:}\ {\it {}For any positive integer $n$ there are 
constants $K$ and $\delta$ depending only on the {\it a-priori} bound on the
second moments of $f$, such that
$$ 2\int_{\IR^3}(-\Delta)^{n} f(v,t)Q_B(f)(v,t)\d  \le K -
(\ell/2) \|{(-\Delta)^{n/2}} f\|_2^2\Eq(5.4)$$
whenever
$$\|f(\cdot,t) - M_{f(\cdot,t)}(\cdot)\|_1 \le \delta\ .$$
where $\ell$ is the constant in \equ(2.3.9).}
\bigskip
\noindent {\bf Proof:}  
It is sufficient to  
note that
$$\eqalign{&2\int_{\IR^3}(-\Delta)^{n} fQ_B(f)\d = \cr
&2\ell\int_{\IR^3}f(-\Delta)^n f\circ f\d - 
2\ell\|(-\Delta)^{n/2}f\|_2^2\le\cr &
2\ell\bigl(\|(-\Delta)^{n/2}f\|_2\|(-\Delta)^{n/2}f\circ f\|_2 -
\|(-\Delta)^{n/2}f\|_2^2\cr}$$
What we need now is control over $\|(-\Delta)^{n/2}f\circ f\|_2$.
This is provided by an inequality from [\rcite{CGT}], where it is shown
that for any $\gamma>0$ there are constant $K$ and $\delta >0$ such
that 
$$\|(-\Delta)^{n/2}f\circ f\|_2^2 \le K  + \gamma \|(-\Delta)^{n/2} 
f\|_2^2.\Eq(5.6)$$ whenver
$$\|f - M_f\|_1 \le \delta\ .$$

This inequality is proved in [\rcite{CGT}] assuming that $f$ has zero
mean and unit variance, and under these conditions, the constants $K$
and $\delta$ are universal. Scaling the inequality, we have it holding
with constants $K$ and $\delta$ depending only on the second moments
of $f$.  (The inequality is applied in [\rcite{CGT}] to get stong
exponential convergence estimates for the spatially homogeneous
Boltlzmann equation with physically realistic constants in the
bounds.)

With this inequality, we need only take $\gamma = 1/2$. \ $\square$
\bigskip
\noindent{\bf Proof of Theorem 5.1:} We begin with the proof of
\equ(5.1).

To put all of the lemmas together, let $\tau := (1/\epsilon)t$
as before, and put
$x(\tau) := \|f(\cdot,\tau)\|_2^2$. Combining Lemmas 5.2 and 5.3, we 
get in the BGK case that
$$\dot x(\tau) \le \epsilon K + K - x(\tau)$$
and any solution of this differential inequality satisfies
$x(\tau) \le x(0) + (1+ \epsilon)K$ for all $\tau$. This establishes 
\equ(5.1) in the BGK case.
The proof of \equ(5.2) in the BGK case is entirely analogous.

To handle the Maxwellian collision kernel case, first define
$$\bar t= \inf \{t \hbox{ such
that } \gamma \|f(\cdot,t)-M_{f(\cdot,t)}\|_1^2\ \ge\ \delta/2 \}
\Eq(5.7)$$ 
where $\delta$ is the universal constant from Lemma 5.4.
We take our initial condition small enough that $\bar t > 0$. Then, for
all $t \le \bar t$, we have the following differential ineqaulity by
combining Lemmas 5.2 and 5.4:
$$\dot x(\tau) \le \epsilon K + K - \ell x(\tau)$$ 
and any solution of this differential inequality satisfies
$x(\tau) \le x(0) + (1+ \epsilon)K/\ell$ for all $\tau \le (1/\epsilon)
 \bar t$.

It only remains to  show that actually $\bar t = +\infty$. We shall do 
this in the next section using an entropy inequality. The entropy
inequlity  requires the {\it a-priori} smoothness bounds from Theorem
5.1, and shows  that a long as they hold; i.e., as long as $t \le \bar
t$, 
we have an upper bound of the form
$$\|f(\cdot,t)-M_{f(\cdot,t)}\|_1 \le \delta_{\rm entropy}(\epsilon)\ .
$$ All we have to do now is to take $\epsilon$ so small that
$$\delta_{\rm entropy}(\epsilon) \le \delta/4$$
and then it is clear from the definition of $\bar t$ that $\bar t = 
\infty$. Thus, borrowing the entropy bound from the next section,
\equ(5.1) is established for $Q_B$ for all $\epsilon$ sufficiently
small, and all initial data sufficently close to a Maxwellian. Again,
\equ(5.2) for $Q_B$ is handled in an entirely analogous way. \ 
$\square$

\bigskip
\noindent{\bf 6. Entropy bounds}
\bigskip
\numsec= 6
\numfor= 1 
\bigskip
 
Let $h(\rho_1|\rho_2)$ denote the relative entropy of two probability
densities $\rho_1$ and $\rho_2$ on $\IR^3$:
$$h(\rho_1|\rho_2) = \int_{\IR^3} \rho_1 \biggl({\rho_1\over \rho_2}
\biggr)\ln
\biggl({\rho_1\over \rho_2}\biggr)\d\ .$$
  
Here we are primarily interested in bounds on $h(f|M_f)$. It will be
convenient however, to first obtain bounds on $h(f|M)$, and to then
relate the two relative entropies. We do this in the next two lemmas.
 
\bigskip
 
\noindent{\bf Lemma 6.1} {\it Let $f$ be any solution of \equ(2.1)
with 
$$\langle |v|^4\rangle_0 \le C \qquad{\rm and}\qquad
\|\nabla f\|_2 < C$$
for all $t\ge 0$. 
Then there is
a constant $K$ depending only on $C$ such that
$${{\rm d}\over {\rm d}t}h(f(\cdot,t)|M) \le K + {1\over \epsilon}
\int_{\IR^3}f(\cdot,t)Q(\cdot, t))\d\ .\Eq(6.1)$$}
\bigskip
 
\noindent{\bf Proof:} Differentiating, we have
$$\eqalign{&{{\rm d}\over {\rm d}t}h(f(\cdot,t)|M) =\cr
&\int_{\IR^3}{\partial f\over \partial t}\bigl(\ln f - \ln M\bigr)\d =
\cr &\int_{\IR^3}\bigl(-E\cdot\nabla f + Lf + \epsilon^{-1}Q(f)\bigr)
\bigl(\ln f - \ln M\bigr)\d\cr}$$
 
Now, integration by parts reveals that 
$$\int_{\IR^3}Lf\bigl(\ln f - \ln M\bigr)\d \le 0$$
and since $\ln M$ is linear in $|v|^2$, 
$$-\int_{\IR^3}\ln M Q(f) = 0\ .$$
{}Finally,
$$\eqalign{&\int_{\IR^3}\bigl(E\cdot\nabla f\bigr)\bigl(\ln f -  \ln
M\bigr)\d =\cr &\int_{\IR^3}\bigl(E\cdot\nabla f\bigr)\ln f\d -  
\int_{\IR^3}\bigl(E\cdot\nabla f\bigr)\ln M\d =\cr
&E\cdot\langle v\rangle_t\cr}$$
Using Jensen's inequality to bound $|\langle v\rangle_t|$ in terms
of the uniformly bounded $\vt$, we have the assreted result.\quad 
$\square$
 
\bigskip
 
\noindent{\bf Lemma 6.2} {\it Let $f$ be any solution of (2.1)
with 
$$\langle |v|^4\rangle_0 \le C \qquad{\rm and}\qquad
\|\nabla f\|_2 < C$$
for all $t \le T$, some $T>0$. Then there is a constant $K$ depending
only on $C$ so that such that
$${{\rm d}\over {\rm d}t}h(f(\cdot,t)|M_{f(\cdot,t)}) \le K +  {1\over
\epsilon}
\int_{\IR^3}f(\cdot,t)Q(\cdot, t))\d\ .\Eq(6.2)$$
for all $t \le T$.}
\bigskip

\noindent{\bf Proof:} By the definitions, we have
$$\eqalign{& h(f(\cdot,t)|M_{f(\cdot,t)}) - h(f(\cdot,t)|M) =\cr
&\langle \ln M\rangle_t - \langle \ln M_{f(\cdot,t)}\rangle_t = \cr
&(3/2)\ln\langle |v-\langle v\rangle_t|^2\rangle_t + (3/2) - 
(1/2)\vt\cr}$$  Thus, 
$$\eqalign{&{{\rm d}\over {\rm d}t}h(f(\cdot,t)|M_{f(\cdot,t)}) =\cr
&{{\rm d}\over {\rm d}t}h(f(\cdot,t)|M) +\cr
&(3/2){{\rm d}\over {\rm d}t}\ln \langle |v-\langle v\rangle_t|^2
\rangle_t - (1/2){{\rm d}\over {\rm d}t}\vf\cr}$$
 
Computing further with the logarithmic derivative term,
 
$$\eqalign{&{{\rm d}\over {\rm d}t}\ln\langle |v-\langle v\rangle_t|^2
\rangle_t =\cr &\bigl(\langle |v-\langle
v\rangle_t|^2\rangle_t\bigr)^{-1} {{\rm d}\over {\rm d}t}\bigl(\langle
|v|^2\rangle_t  -\langle v\rangle_t^2\bigr)\cr}$$
 
To control this term, we need an upper bound on 
$\bigl(\langle |v-\langle v\rangle_t|^2\rangle_t\bigr)^{-1}$
But since $\langle |v-\langle v\rangle_t|^2\rangle_t\ge C\|\nabla
f\|_2^{-4/5} \ge K$ by Lemma 4.2, we have on  application of Lemma 5.1 
that
$${{\rm d}\over {\rm d}t}\ln\langle |v-\langle v\rangle_t|^2\rangle_t 
\le K{{\rm d}\over {\rm d}t}\bigl(\langle |v|^2\rangle_t 
-\langle v\rangle_t^2\bigr)$$
The lemmas of Section 3 provide uniform bounds on this last term. and
therefore provide uniform bounds on  the derivative of
$h(f(\cdot,t)|M_{f(\cdot,t)}) - h(f(\cdot,t)|M)$. The result now 
follows from the previous Lemma.\quad$\square$
\bigskip
 
\noindent{\bf Lemma 6.3}\ {\it Let $Q=Q_{BGK}$. {}For any density $f$ 
with finite second moments,
$$\int_{\IR^3}\ln fQ(f)  \le -h(f|M_f)\Eq(6.3)$$}
\bigskip
\noindent{\bf Proof:} Let $S(f) = -\int_{\IR^3}f\ln f\d$ denote the 
entropy of 
$f$. {}For any density $f$ and for any $\tau\in [0,1]$, put $f(\tau) 
=(1-\tau)f  +\tau M_f$. We have
$${{\rm d}\over {\rm d}\tau}f(\tau)=M_f-f= Q(f)$$
and $f(0) = f$. Then
$$\int_{\IR^3}\ln f Q(f)\d = -{{\rm d}\over {\rm
d}\tau }S(f(\tau)).$$ 
However, the entropy functional is concave, so that
$$S(f(\tau)) \ge (1-\tau)S(f) + \tau S(M_f)$$
and hence
$${S(f(\tau))-S(f(0))\over \tau} \ge S(M_f) - S(f).$$
Since $M_f$ and $f$ share the same hydrodynamic moments, 
$S(M_f)- S(f) = h(f|M_f)$\quad$\square$
\bigskip
 
\noindent{\bf Lemma 6.4} {\it Let $Q=Q_{BGK}$ and  $f$ be any solution 
of
\equ(2.1) 
with 
$$\langle |v|^4\rangle_0 \le C \qquad{\rm and}\qquad
\|\nabla f\|_2 < C$$
for all $t$ Then there is
a constant $K$ depending only on $C$ so that
$${{\rm d}\over {\rm d}t}h(f(\cdot,t)|M_{f(\cdot,t)}) \le K -  {1\over
\epsilon} h(f(\cdot,t)|M_{f(\cdot,t)})\ .\Eq(6.4)$$}
\bigskip
 
\noindent{\bf Proof:} This follows immediately upon combining the last 
three lemmas.\quad$\square$

In the case of the Boltzmann collision kernel, Lemma 6.3 is replaced by the
following proposition proved in [\rcite{CC2}]:
\bigskip
\noindent{\bf Proposition 6.5} {\it {}For all $C>0$, there is a positive 
function
$\Phi_C(r)$ strictly increasing in $r$, such that 
for all densities $f$ with
$$\int_{\IR^3}|v|^4f(v)\d \le C\qquad{\rm and}\qquad \|\nabla f\|_2 \le 
C$$ 
$$\int_{\IR^3}\ln fQ_B(f)  \le -\Phi_C[h(f|M_f)].\Eq(6.5)$$}
Consequently, for $Q=Q_B$, \equ(6.4) is replaced by
$${{\rm d}\over {\rm d}t}h(f(\cdot,t)|M_{f(\cdot,t)}) \le K -  {1\over
\epsilon}\Phi_C[ h(f(\cdot,t)|M_{f(\cdot,t)})]\ .\Eq(6.6)$$

Now if $f(\cdot,0) = M_{f(\cdot,0)}$, so that $h(f(\cdot,0)|M_{f(\cdot,
0)} = 0$, then it is evident from this differential inequality that
$$\Phi_C[ h(f(\cdot,t)|M_{f(\cdot,t)})] \le \epsilon K$$
for all $t\ge 0$.
This together with Kullback's inequality
$$\|f(\cdot,t) -M_{f(\cdot,t)} \|_1^2 \le
2h(f(\cdot,t)|M_{f(\cdot,t)}) \Eq(6.8)$$ 
clearly implies that there is
a function $\delta_{\rm entropy}(\epsilon)$, decreasing to zero with
$\epsilon$ so that
$$\|f(\cdot,t) -M_{f(\cdot,t)} \|_1 \le \delta_{\rm entropy}(\epsilon)
$$ for all $t \le \bar t$ --the times for which we know that $f(\cdot,
t)$ will satisfy the bounds in the hypothesis of Proposition 6.5. But
as explained at the end of the proof of Theorem 5.1, this is enough to
show that $\bar t = \infty$ for all sufficiently small $\epsilon$.
Thus, the entropy bound provides the information needed to complete the
proof of Theorem 5.1 as claimed, and moreover, \equ(5.7) hold globaly 
in time. 

\bigskip
\noindent{\bf 7. Proof of Theorems 2.1 and 2.2}
\bigskip
\numsec= 7
\numfor= 1 
\bigskip

\noindent{\bf Proof of Theorem 2.1}\ Let us consider first the case
$Q=Q_{BGK}$. Since Maxwellian initial data satisfies the hypotheses of 
Lemma 6.4, we have that \equ(6.4) holds for the solution
$f(\cdot,t)$ of \equ(2.1) under consideration in Theorem 2.1. But then
$${{\rm d}\over {\rm d}t}\bigl(e^{t/\epsilon}h(f(\cdot,t)|M_{f(\cdot,t)})\bigr)
\le K e^{t/\epsilon}\ .$$
Since by hypothesis, $h(f(\cdot,0)|M_{f(\cdot,0)}) = 0$, we have
$$h(f(\cdot,t)|M_{f(\cdot,t)}) \le \epsilon K$$
for all $t\ge 0$. This together with Kullback's inequality \equ(6.8)
yields the first inequality asserted in Theorem 2.1, with
$\delta_1(\e)=\e^{1/2}$.

To obtain the second, note that since $f(\cdot,0)$ is Maxwellian, there is
a bound on $\|\Delta f\|_2$ depending only on $\langle |v|^2\rangle_0$. 
Inequality \equ(5.2) of Lemma 5.1 now fives us a uniform bound on
$\|\Delta f(\cdot,t)\|_2$. Combining this with the interpolation 
inequality
\equ(4.1) finally yields the second inequality of Theroem 2.1 with
$\delta_2(\e)=\e^{1/7}$ for $Q=Q_{BGK}$.

The proof for $Q=Q_B$ is only slightly more involved, but in fact
we have already given the proof of the first part of Theorem 2.1 in our 
``back and forth'' proof of the smoothness bounds and entropy bounds
for this case. As observed at the end of Section 6,
$$\|f(\cdot,t) -M_{f(\cdot,t)} \|_1 \le \delta_{\rm entropy}(\epsilon)
$$ for all $t\ge 0$, and $\delta_{\rm entropy}(\epsilon)$ does decrease
to 0 with $\epsilon$ as required. The second part follows in an
entriely similar way.\ 
$\square$

\bigskip
 
\noindent{\bf Proof of Theorem 2.2}\ This  will follow from Theorem 
2.1, and it is now no longer necessary to separate the cases $Q =
Q_{BGK}$ and $Q = Q_B$.

As we have computed in the proof of
Lemma 3.1,
$$\eqalign{\phantom{.}&{{\rm d}\over {\rm d}t}\langle |v|^2 \rangle_t 
=\cr
\phantom{.}&-2E\cdot\langle v \rangle_t + 6\langle D \rangle_t
+2(v\cdot\nabla D)\langle (v\cdot\nabla D) \rangle_t 
-2\langle D|v|^2 \rangle_t\cr}$$
If we replace the density $f(\cdot,t)$ everywhere on the right by
$M_{f(\cdot,t)}$, by definition we obtain the function $G(u(t),e(t))$
where $u(t)$ and $e(t)$ are the moments of $f(\cdot,t)$ figuring in 
Theorem 2.2. The error we make has to be estimated term by term. The
least trivial of these terms concerns the contribution  from $\langle
D|v|^2
\rangle_t$, and is estimated as follows:
$$|\langle D|v|^2 \rangle_t -
\int_{\IR^3}D(v)|v|^2M_{f(\cdot,t)}(v)\d | \le$$
$$\int_{\IR^3}D(v)|v|^2|f(v,t) - M_{f(\cdot,t)}(v)|\d\ \le$$
$$\int_{|v|\le R}D(v)|v|^2|f(v,t) - M_{f(\cdot,t)}(v)|\d\ +$$
$$\int_{|v|\ge R}D(v)|v|^2|f(v,t) - M_{f(\cdot,t)}(v)|\d\ \le$$
$$C\biggl(R^2\|f(\cdot,t) - M_{f(\cdot,t)}\|_1 + R^{-2}
\int_{\IR^3}|v|^4|f(v,t) - M_{f(\cdot,t)}(v)|\d\biggr)\le$$
$$C\bigl(R^2\|f(\cdot,t) - M_{f(\cdot,t)}\|_1 + 
R^{-2}K\bigl(\vt + 1\bigr)\bigr)$$
Optimizing over $R$ now yields the result
$$|\langle D|v|^2 \rangle_t - \int_{\IR^3}D(v)|v|^2M_{f(\cdot,t)}(v)
\d |\le K\|f(\cdot,t) - M_{f(\cdot,t)}\|_1^{1/2}$$
Theorem 2.1 yields a bound of the size $\delta_1(\e)^{1/2}$.
The other error terms in the time derivatives of $\vt$ and
$\langle v\rangle_t$ are bounded by a direct application of  Theorem
2.1  (and hence yield errors of order $\delta_1(\e)$ instead of
$\delta_1(\e)^{1/2}$). \quad$\square$

\bigskip
\noindent{\bf 8. Stationary solutions}
\bigskip
\numsec= 8
\numfor= 1 
\bigskip

In this section we discuss stationary solutions of \equ(2.1), and
prove Theorem 2.4. The stationary solutions of \equ(2.1) are the
positive normalized solutions of
$$\e{\cal L} f+ Q(f)=0. \Eq(8.1)$$
The existence of such solutions relies on a simple fixed point argument.
Recalling the expression of $Q(f)$, we put
$$J(f)=\cases{ M_f \quad \hbox{\rm in the BGK case },\cr
f\circ f \quad\hbox{\rm in the Boltzmann case }.}
$$
Equation \equ(8.1) can be rewritten as
$$f=J(f)+ {\e\over \ell} {\cal L} f,\Eq(8.2)$$
where $\ell=1$ in the BGK case. The explicit form of the linear 
operator ${\cal L}$ implies that, for $\e$ sufficiently small, the operator
$${\cal G} = (1-{\e\over \ell} {\cal L})^{-1},\Eq(8.3)$$ 
is positivity and normalization preserving on $L_1(\IR^3)$. (That is, it is a
Markovian operator). Moreover, it is clear from the smoothing
properties of of ${\cal G}$ that it is compact on $L_1(\IR^3)$. Also,
$J$ is a positivity and normalization preserving map of $L_1(\IR^3)$
into itself. Then since we can rewrite (8.2) as
$$f={\cal G}J(f).\Eq(8.4)$$ we see that the solutions of \equ(8.1)
that we seek are the fixed points of the map $f\mapsto{\cal G}J(f)$.
The properties of this map listed above prove the existence of fixed
points (see [\rcite{Kra}]).  Let us denote by $f_*$ one of them. It is
then easy to check that it solves
\equ(8.1) pointwise. 

Next we note that the same arguments used in previous sections imply that, if
$f$ solves\equ(8.1) then
$$\|f_*-M_{f_*}\|\le\delta_{\rm entropy}(\epsilon).$$
This concludes the first part of Theorem 2.4. 

Everything done so far would apply in the Boltzmann case as well as
the BGK case. What we don't know at this point is: Are there fixed
points in {\it each} of the stable neighborhoods, and if so is there
{\it exactly one} in each stable neighborhood.

These questions can be positively answered in the BGK case in a
simplre way. It seems likely that one could also provide a positive
answer for at least the first of them for the Boltzmann kernel, but we
have not done more than skecth a lengthly argument, and so will
confine ourselve to the BGK case.

In fact, since ${\cal G}=I+\e{\cal L}{\cal G}$ with $I$ the identity map and
$J(f)=M_f$, we can write 
\equ(8.4) as
$$f=M_\e+\e{\cal L} {\cal G} M_\e.\Eq(8.5)$$
Note that the right hand side of \equ(8.5) depends only on $u$ and $e$, the
first and second moments of $f$. If we multiply \equ(8.5) by $v$ or by
$v^2$ and integrate, we get
$$\eqalign{ F_\e(u,e)=0,\cr G_\e(u,e)=0, }\Eq(8.6)$$ because $f$ and
$M_f$ have the same first two moments. The functions $F_e$ and $G_e$
are quite complicated, but for $\e=0$ they reduce to the functions $F$
and $G$ in the right hand side of \equ(2.5.3).  Then, by Proposition
2.3 we know there are solutions $(u_*,e_*)$ to \equ(8.6) for
$\e=0$. Moreover the differential of the map $(u,e)\to
(F_\e(u,e),G(u,e))$ has eigenvalues with non vanishing real part, in
$\e=0$ and $(u,e)=(u_*,e_*)$, when $E$ is in the appropriate
range. Therefore, by the implicit function theorem, for $\e$
sufficiently small, we have a unique solution $(u_\e, e_\e)$ in a
neighborough of $(u_*,e_*)$ to \equ(8.6).  Let $M_\e$ be the
Maxwellian with moments $(u_\e, e_\e)$. Then it is easy to check that
$$f=M_\e+{\cal L}{\cal G}M_\e$$
is solution to \equ(8.1). This concludes Theorem 2.4.

\bigskip
\noindent{\bf 9. Long time behavior.}
\bigskip
\numsec= 9
\numfor= 1 
\bigskip

It is natural to ask whether the stationary solution are the
asymptotic limits as $t\to+\infty$ of the evolution starting in
appropriate neighborhoods of the fixed point Maxwellian. To this we
have only a partial answer even in the BGK case:

{\bf Proposition 9.1} {\it Choose a stable fixed point $(u^*,e^*)$ of 
\equ(2.5.3) and let $f_*$ be a stationary solution of \equ(2.1) in the
neighborhood of $M_{(u^*,e^*)}$. Assume that the solution $f_t$ of the
time dependent problem starting near $M_{(u^*,e^*)}$ has moments
$u(t)$ and $e(t)$ converging to $u_*$ and $e^*$ respectively.  Then
$$\lim_{t\to+\infty} ||f-f_*||_2=0.$$
}

Unfortunately we do not have enough control on the time behavior of
the solution to check the convergence of the moments. We expect
however such convergence and this can be proven for a modified model
where we consider, instead of a diffusion coefficient $D$ depending on
the velocity, one depending only on the average $e$.  A
straightforward calculation then shows that one gets closed equations
for the first two moments and the long time asymptotics is easily
obtained.  In this case all our results still apply and the conditions
of Proposition 2.5 are fulfilled.

{}First we prove Proposition 2.5. Let $f_*$ be a fixed point and $f$
the distribution at time $t$. Calculating as in previous sections, we
have:$$\eqalign {
&{1\over 2}{{\rm d}\over {\rm d}t}\|f-f_*\|_2^2= \int_{\IR^3}\d
(f(v)-f_*(v))[{\cal L}f(v) +{1\over \e}Q(f)(v)]\cr
&= \int_{\IR^3}\d (f(v)-f_*(v))[{\cal L}(f(v)-f_*(v)) +{1\over
\e}(Q(f)(v)-Q(f_*)(v))]\cr
&=\int_{\IR^3}\d (f(v)-f_*(v)){\cal L}(f(v)-f_*(v))\cr
&+{1\over \e}\int_{\IR^3}\d
(f(v)-f_*(v))(M_f(v)-M_{f_*}(v))-{1\over\e}\|f-f_*\|_2^2. }$$
It is easy to chech that
$$\int_{\IR^3}\d (f(v)-f_*(v)) L_2(f(v)-f_*(v))\le 0,$$
$$\int_{\IR^3}\d(f(v)-f_*(v))E\cdot\n_v(f(v)-f_*(v))=0.$$
On the other hand
$$\eqalign
{
&\int_{\IR^3}\d (f(v)-f_*(v)) L_1(f(v)-f_*(v))\cr
&=-\int_{\IR^3}\d \n_v (f(v)-f_*(v)) D(v)M(v)\n_v\left({f(v)-
f_*(v)\over M}\right)\cr
&=-\int_{\IR^3}\d |\n_v (f(v)-f_*(v))|^2 D(v) - \int_{\IR^3}\d v\cdot 
\n_v (f(v)-f_*(v)) D(v)(f(v)-f_*(v))\cr
&=-\int_{\IR^3}\d |\n_v (f(v)-f_*(v))|^2 D(v) -  {1\over
2}\int_{\IR^3}\d v\cdot
\n_v (f(v)-f_*(v))^2 D(v)\cr
&=-\int_{\IR^3}\d |\n_v (f(v)-f_*(v))|^2 D(v) +  {1\over
2}\int_{\IR^3}\d  (f(v)-f_*(v))^2\n_v\cdot(v D(v))\cr
&=-\int_{\IR^3}\d |\n_v (f(v)-f_*(v))|^2 D(v) +  {1\over
2}\int_{\IR^3}\d  (f(v)-f_*(v))^2v\cdot \n_v D(v))\cr
&+{3\over 2}\int_{\IR^3}\d (f(v)-f_*(v))^2 D(v)\cr
&\le {3\over 2}\int_{\IR^3}\d (f(v)-f_*(v))^2D(v)\cr
}
$$
The last ineqality is consequence of the \equ(3.3).
Since 
$$\int_{\IR^3}\d (f(v)-f_*(v))(M_f(v)-M_{f_*}(v))\le  {1\over 2}
\|f-f_*\|^2_2 +{1\over 2} \|M_f -M_{f_*}\|^2_2,$$
we conclude that 
$${{\rm d}\over {\rm d}t}\|f-f_*\|_2^2\le {1\over \e}  \|M_f
-M_{f_*}\|^2_2 -[{1\over \e}-3(a+c)]\|f-f_*\|_2^2.$$
With $x(t)=\|f-f_*\|_2^2$ and $a= \|M_f-M_{f_*}\|^2_2$ we have,  for
$\e$ sufficiently small:
$$\dot x + c x\le a$$
and hence
$$x(t)\le x_0{\rm e}^{-{t/e}}+\int_0^td\/s {\rm e}^{-{(t-s)/e}a(s)}.$$
This implies $x(t)\to 0$ as $t\to \infty$, provided that $a(t)\to 0$ 
as $t\to
\infty$. This concludes Proposition 2.5.

The convergence of $M_f$ to $M_{f_*}$ is not easy to get for 
\equ(2.1). A simple answer is obtained if one replaces the operator
$L_1$ given by
\equ(2.2) with
$$L_1f(v)=D(e)\nabla\cdot\biggl(M(v)\nabla\/f\biggl({f(v)\over M(v)}
\biggr)
\biggr),\Eq(9.2)$$
This model is much simpler than the one already considered, but still
has a non trivial behavior on the hydrodynamical sale.  In particular,
it is easy to see that, if we write the equations for $u$ and $e$, we
get a closed system in those variables. Namely,
\equ(2.5.3),\equ(2.5.5) and \equ(2.5.7) are replaced by
$${{\rm d}\over {\rm d}t}\left(\matrix{u(t)\cr
e(t)\cr}\right) = \left(\matrix{F(u(t), e(t))\cr
G(u(t), e(t))\cr}\right),\Eq(9.5.3)$$
the functions $F$ and $G$ being given explicitly by 
$$
F(u,e) = E - u[\nu +D(e)], 
\Eq(9.5.5)
$$
$$
G(u,e) = Eu +3 D(e)-2eD(e). \Eq(9.5.7)
$$
Equation \equ(9.5.3) is exact for this model independently of $\e$ and
for suitable choices of the functions $D(e)$ has several critical
points for $E$ in an appropriate range. Moreover the asymptotic
behavior for large times is easy to establish. All our results apply
to this model without substantial changes. In particular, in this case
we can use Proposition 2.5 to obtain the convergence to stationary
solutions for large times.

\bigskip

\noindent{\bf Acknowledgements}

This research was supported in part by NSF Grant DMS--920--7703, CNR--GNFM
and MURST, and AFOSR Grant AF--92--J--0015.

\bigskip

\bigskip
\centerline{\bf References}
\vskip.3cm

\item{[\rtag{N}]} T. Nishida,  {\it  Fluid
dynamical limit of the nonlinear Boltzmann equation to
the level of the compressible Euler equation }, 
Commun. Math. Phys. {\bf 61},  119-148, (1978).

\item{[\rtag{Ca}]}  R. E. Caflisch, {\it The
fluid dynamic limit of the nonlinear Boltzmann equation},
Commun. on Pure and Applied Math. {\bf 33},  
651-666, (1980). 

\item{[\rtag{U}]} S. Ukai, K. Asano {\it The Euler limit
and initial layer of the  nonlinear Boltzmann equation}, 
Hokkaido Math. J. {\bf 12},  303-324, (1983).

\item{[\rtag {DEL}]}
A. De Masi, R. Esposito, and J. L. Lebowitz, 
{\it Incompressible Navier-Stokes and Euler limits of the 
Boltzmann equation},  
Commun. Pure Appl. Math., {\bf 42},  1189--1214, (1989).

\item{[\rtag{ELM1}]} 
R. Esposito,J. L.Lebowitz and R.  Marra
{\it Hydrodynamical limit of the  Stationary Boltzmann Equation   in
a Slab},
Commun. Math. Phys. {\bf 160}, 49--80 (1994).

\item{[\rtag{ELM2}]}  
R. Esposito,J. L.Lebowitz and R.  Marra
{\it Navier-Stokes behavior  of
stationary  solutions of the  Boltzmann Equation}.
Jour.Stat.Phys. {\bf 78},389--412 (1995).

\item{[\rtag{R}]}  
\ A. Rokhlenko, Phys. Rev.A, {\bf 43}, 4438
(1991);  A. Rokhlenko and J.L. Lebowitz, Phys. Fluids B:  Plasma
Physics, {\bf 5}, 1766 (1993).

\item{[\rtag{B}]}
\ R. Balescu, {\it Transport Processes in
Plasmas}, North-Holland (1988).

\item{[\rtag{F}]}  \ R.N. Franklin, {\it Plasma Phenomena in Gas
Discharges}, Clarendon Press, Oxford, (1976):  N.J. Carron, Phys. Rev.
{\bf A45}, 2499 (1992).

\item{[\rtag{CC1}]}\ E. A. Carlen, M. C. Carvalho,
{\it Strict Entropy Production Bounds and Stability of the Rate of
Convergence to Equilibrium for the Boltzmann Equation},
Jour. Stat. Phys. {\bf 67}, 575--608 (1992).

\item{[\rtag{CC2}]} \ E. A. Carlen, M. C. Carvalho, 
{\it Entropy Production Estimates for Boltzmann Equations with
Physically Realistic Collision Kernels},
Jour. Stat. Phys. {\bf 74}, 743--782 (1994).

\item{[\rtag{CELMR}]}\ 
E. Carlen, R.Esposito, J.L.Lebowitz, R.Marra  and A.Rokhlenko, 
{\it 
Nonunique Stationary States in Driven Collisional  Systems with
Application to Plasmas},\
Phys. Rev E {\bf 52}, 40--43, (1995).

\item{[\rtag{CGT}]} \ E. A. Carlen, E. Gabetta, G. Toscani, {\it Propagation of
Smoothness in Velocities and Strong Exponential Convergence for
Maxwellian Molecules} Georgia Tech and Pavia Preprint.

\item{[\rtag{Kra}]} \ Krasnosel'ski M., {\it Topological Methods in
the Theory of Nonlinear Integral Equations} MacMillan Press, New York
(1964) \end